\documentclass[aps,prl,superscriptaddress,reprint,twocolumn]{revtex4-2}

\usepackage{mathrsfs}  
\usepackage{amssymb}
\usepackage{color} 
\usepackage{ulem}
\usepackage{mathptmx}
\usepackage{enumitem}
\usepackage{amsmath}    
\usepackage{graphicx}   
\usepackage{verbatim}   
\usepackage{hyperref}   
\DeclareMathOperator{\f}{f}

\begin{document}
\title{Too fast to grow: Dynamics of pendant drops sliding on a thin film}
\author{Etienne Jambon-Puillet}
\affiliation{Department of Chemical and Biological Engineering, Princeton University,Princeton, New-Jersey 08540, USA}
\author{Pier Giuseppe Ledda}
\affiliation{Laboratory of Fluid Mechanics and Instabilities, École Polytechnique Fédérale de Lausanne, Lausanne, CH-1015, Switzerland}
\author{François Gallaire}
\affiliation{Laboratory of Fluid Mechanics and Instabilities, École Polytechnique Fédérale de Lausanne, Lausanne, CH-1015, Switzerland}
\author{P.-T. Brun}
\affiliation{Department of Chemical and Biological Engineering, Princeton University,Princeton, New-Jersey 08540, USA}

\date{\today}

\begin{abstract}
Pendant drops suspended on the underside of a wet substrate are known to accumulate fluid from the surrounding thin liquid film, a process that often results in dripping. The growth of such drops is hastened by their ability to translate over an otherwise uniform horizontal film. Here we show that this scenario is surprisingly reversed when the substrate is slightly tilted ($\approx 2$ deg) ; drops become too fast to grow and shrink over the course of their motion. Combining experiments and numerical simulations, we rationalize the transition between the conventional growth regime and the previously unknown decay regime we report. Using an analytical treatment of the Landau-Levich meniscus that connects the drop to the film, we quantitatively predict the drop dynamics in the two flow regimes and the value of the critical inclination angle where the transition between them occurs.

\end{abstract}

\maketitle

\begin{figure}[b]
    \begin{center}
        \includegraphics[width=0.9\columnwidth]{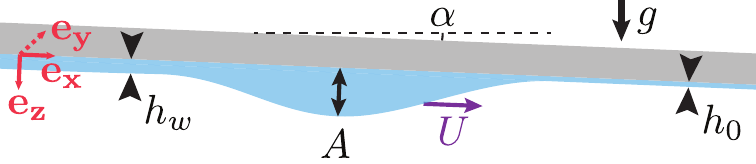}
        \caption{Schematic of a pendant drop of amplitude $A$ sliding with velocity $U$ under a substrate pre-wetted with a film of thickness $h_0$ and inclined by an angle $\alpha$. 
        } 
        \label{fg:fig1}
    \end{center}
\end{figure}

Anyone who has applied paint to a ceiling knows that thin liquid coatings can spontaneously destabilize and accumulate into an array of pendant drops~\cite{Yiantsios:1989,Fermigier:1992}. While interfacial instabilities can be harnessed to build structures akin to geomorphic patterns~\cite{Weitz:2008,Marthelot:2018,Jambon:2021,Ribe:2007,camporeale2012hydrodynamic,Dutta:2016}, the Rayleigh-Taylor instability in films is more commonly seen as undesirable, e.g. jeopardizing the uniformity of coatings~\cite{Weinstein:2004}. 
Worse, as they grow, instability-mediated drops can drip and pollute the space underneath, with potentially severe consequences for  engineering constructs~\cite{Kaita:2010,vanEden:2017}.
As such, the Rayleigh-Taylor instability in thin viscous films has been extensively studied~\cite{Yiantsios:1989,Fermigier:1992,Limat:1992,Yoshikawa:2019,Lerisson:2020} and diverse strategies have been proposed to prevent the formation of drops~\cite{Burgess:2001,Lapuerta:2001,Alexeev:2007,Cimpeanu:2014,Trinh:2014,Brun:2015,Balestra:2018}. 
Linear stability analysis for this class of problems is therefore well established, while insights in the drop patterns formed by the instability have been provided using weakly nonlinear developments~\cite{Fermigier:1992}. Yet, our understanding of fully formed pendant drops and their transition to dripping remains sparse~\cite{Lister:2010} owing to the difficulties of modeling the fully non-linear long-term dynamics. In this Letter, we focus on a single pendant drop (see Fig.~\ref{fg:fig1}), a problem that remains analytically tractable while retaining a rich physics. 

Pendant drops under uniformly coated films are capable of steady translation, even in the theoretical limit of a perfectly horizontal substrate~\cite{Lister:2010}. Over the course of their trajectory, these drops accumulate more fluid from the surrounding thin film than if they were stationary~\cite{Lister:2010}, thereby reaching the critical size leading to dripping~\cite{Marthelot:2018} faster than immobile drops. Here, using experiments, numerical simulations and theory, we show that increasing the drop velocity by slightly tilting the substrate surprisingly prevents dripping. Past a critical inclination, the film left by the drop in its wake is thicker than the one absorbed by the drop in its front. This negative balance depletes the volume of fluid in the drop, which shrinks, thereby avoiding dripping. Through an analysis of the Landau-Levich meniscus at the edge of the drop, we unveil the physics at play in these drops that are too fast to grow and  predict analytically their dynamics and the transition between the two aforementioned flow regimes. 

Our experiment is schematized in Fig.~\ref{fg:fig1}. Silicone oil (density $\rho=971$ kg/m$^3$, surface tension $\gamma=20.3$ mN/m, viscosity $\eta=1.13$ Pa.s) is spin-coated on a flat glass substrate to produce a film of uniform thickness $h_0$ (measured by weighting the sample). The substrate is then flipped and mounted onto a rotating arm while a droplet is applied on the film with a micropipette. The resulting pendant drop has an initial amplitude  $A_0\sim \ell_c$ where $\ell_c=\sqrt{\gamma/(\rho g)}$ denotes the capillary length and $g$ denotes the acceleration of gravity. The substrate is then tilted by an angle $\alpha$ and the dynamics is recorded (see SM section I for details~\footnote{See Supplemental Material at [URL will be inserted by publisher] for experimental and numerical methods, additional experimental and numerical results and more details about the model and its limitations.}). Note that the initial coating is sufficiently thin not to destabilize via the Rayleigh-Taylor instability over the course of our experiment~\cite{Fermigier:1992}. The film thickness is therefore assumed to be uniform and constant far from the drop.

\begin{figure*}[t]
    \begin{center}
        \includegraphics[width=\textwidth]{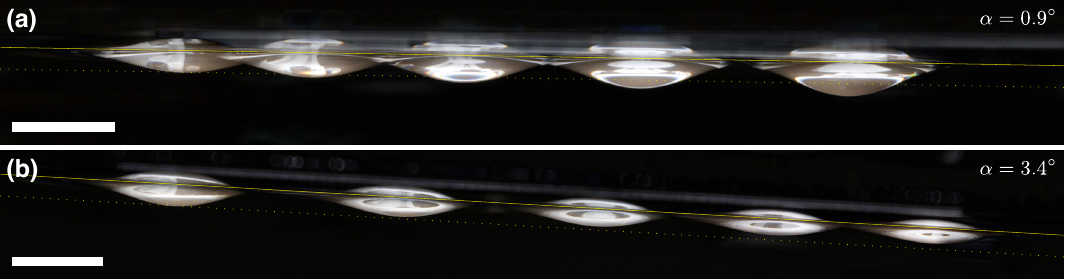}
        \caption{Chronophotographies of two experiments at low \textbf{(a)} and high \textbf{(b)} inclination angles with $h_0\approx 89$ $\mu$m (scale bars are $5$ mm). The interval between pictures is $9$ and $3.75$ min, respectively. The pictures  include the reflexion of the drop  on the substrate. The solid lines indicate the position of the substrate. The dotted lines mark the initial amplitudes of the drops $A_0=\{1.03,1.16\}$ mm. 
        } 
        \label{fg:fig2}
    \end{center}
\end{figure*}

Fig.~\ref{fg:fig2}(a) shows a chronophotography of an experiment performed with a nearly horizontal substrate ($\alpha=0.9^\circ$). As evident from the figure, the drop translates by several times its diameter over the course of the experiment while both the drop speed $U$ and amplitude $A$ increase. In Fig.~\ref{fg:fig2}(b), we show an experiment nearly identical to (a), except for that the inclination angle is slightly higher ($\alpha=3.4^\circ$). As expected, the drop initially moves faster. However, unlike the lower inclination case, it progressively shrinks and decelerates. 
\begin{figure}[tb]
    \begin{center}
        \includegraphics[width=\columnwidth]{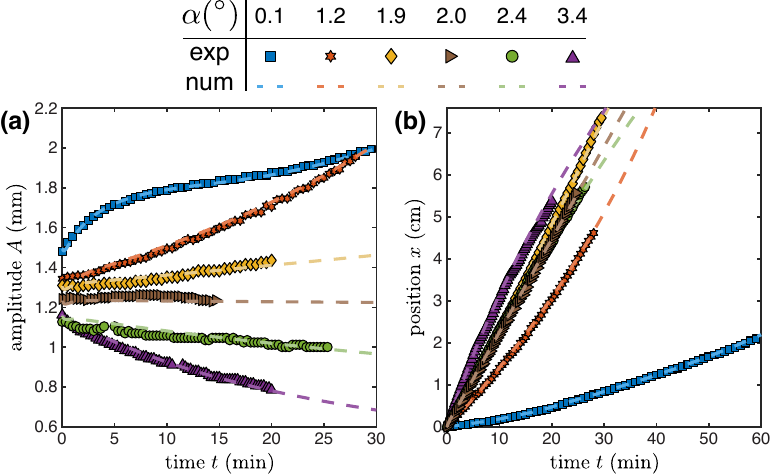}
        \caption{Shown is the amplitude $A(t)$ \textbf{(a)} and position $x(t)$ \textbf{(b)} of drops sliding under a film of thickness $h_0\approx 89$ $\mu$m at different inclination angles $\alpha$ (see legend). Markers indicate experiments and dashed lines indicate numerical simulations. 
        } 
        \label{fg:fig3}
    \end{center}
\end{figure}
In Fig.~\ref{fg:fig3} we plot the amplitude $A(t)$ and position $x(t)$ of drops sliding over films of similar thickness ($h_0\approx 89$ $\mu$m) but with different inclination angles $\alpha$. While the drop accelerates and grows for the three smallest angles, the situation is reversed for the two largest angles. The inclination $\alpha_c\approx 2^\circ$ appears to be the critical angle, $\alpha_c$, where the drop amplitude and speed are constant ($U\approx 2.3$ mm/min). Movie~S1 displays similar observations for $h_0\approx 112$ $\mu$m. Modifying the film thickness changes the value of  $\alpha_c$,  as well as the timescale of the experiment. Increasing the drop initial amplitude $A_0$ appears to speed-up the dynamics but does not change its outcome (see SM section II~\cite{Note1}) . 

We turn to numerical simulations to rationalize these two flow regimes. Owing to the dimensions of the problem we use the lubrication approximation to describe the evolution of the position of the interface $h(x,y,t)$~\cite{Yiantsios:1989}
but retain the full fledged expression of the curvature $\kappa$~\cite{Wilson:1982,Lerisson:2020}. 
In the Cartesian frame aligned with the substrate (see Fig.~\ref{fg:fig1}), we obtain the following dimensionless thin-film equation after rescaling $x$ and $y$ using $\ell_c/\sqrt{\cos\alpha}$, $h$ using the coating thickness far from the drop $h_0$, and $t$ using $\tau=\eta\gamma/\left(h_0^3\rho^2g^2\cos^2\alpha\right)$:
\begin{equation}
\begin{gathered}
\partial_{\,\bar{t}}\bar{h} + \widetilde{\alpha}\bar{h}^2 \partial_{\bar{x}}\bar{h} + (1/3)\mathbf{\bar{\nabla} \cdot} \left[\bar{h}^3\left(\mathbf{\bar{\nabla}}\bar{h} + \mathbf{\bar{\nabla}}\bar{\kappa} \right)\right]=0, \\
\bar{\kappa}=\mathbf{\bar{\nabla} \cdot}\left[\mathbf{\bar{\nabla}}\bar{h}  \middle/\sqrt{1+\left(h_0\sqrt{\cos\alpha}/\ell_c\right)^2 \left(\mathbf{\bar{\nabla}}\bar{h}\right)^2 }\right],
\end{gathered}
\label{eq:lub_adim}
\end{equation}
where a bar indicates rescaled variable. Note that the inclination of the substrate is captured by $\widetilde{\alpha}=\frac{\ell_c \tan\alpha}{h_0\sqrt{\cos\alpha}}\approx \frac{\ell_c \alpha}{h_0}$.

We solve eq.~\eqref{eq:lub_adim} with the finite element software COMSOL on a rectangular domain with periodic boundary conditions and the initial condition $\bar{h}(\bar{x},\bar{y},0)=1+h_d(\bar{x},\bar{y})/h_0$. Here, $h_d(\bar{x},\bar{y})/\ell_c$ is the dimensionless profile of a static pendant drop obtained by integrating the Young-Laplace equation numerically (see SM section I for numerical details~\cite{Note1}). 
In Movie~S2 we report typical numerical results, which appear qualitatively similar to experiments, i.e. predicting growth and ultimately dripping at low inclination angles and the opposite at higher angles. In Fig.~\ref{fg:fig3} we show the evolution of the drop amplitude $A(t)$ and position $x(t)$ obtained numerically with the parameters corresponding to the aforementioned experiments within their uncertainty ($\Delta h_0 = 7$ $\mu$m, $\Delta \alpha = 0.15^\circ$). The agreement between experiments and numerics is favorable thereby validating our simulations. 

\begin{figure*}[tb]
    \begin{center}
        \includegraphics[width=\textwidth]{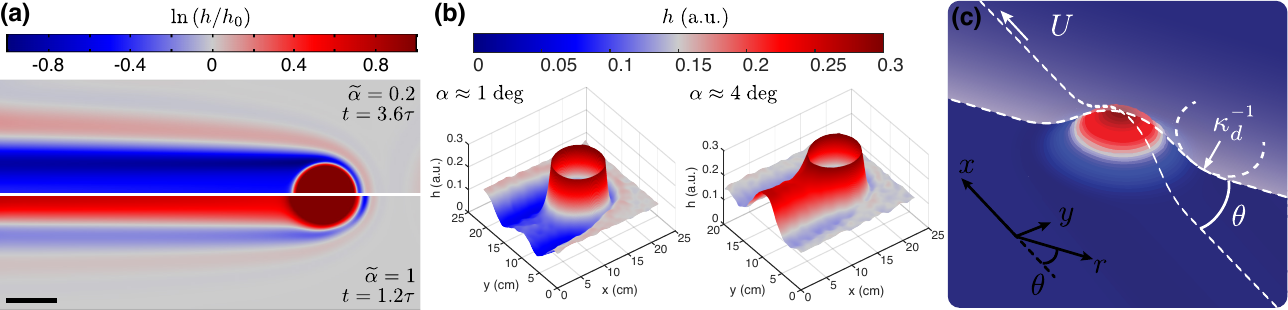}
        \caption{\textbf{(a)} Thickness map for two simulations with identical initial drops ($A_0/\ell_c=1.05$, $h_0/\ell_c=0.04$) but different inclination angles $\widetilde{\alpha}$ (position $x=37.4\ell_c$). Scale bar $5\ell_c$. \textbf{(b)} Thickness profiles inferred from experiments conducted with dyed oil at low and high inclination angles ($A_0/\ell_c\approx 0.8$, $h_0\approx85$ $\mu$m). \textbf{(c)} Three dimensional schematics of the drop introducing the polar coordinates $\{r,\:\theta\}$, and curvature $\kappa_d$.
} 
        \label{fg:fig4}
    \end{center}
\end{figure*}

Leveraging our simulations we investigate the physics setting the value of the critical angle $\alpha_c$.  In Fig.~\ref{fg:fig4}(a) we plot side-by-side the log of the dimensionless film thickness $h(x,y)/h_0$ for two drops, each of which corresponds to a given flow regime. The two situations  only differ in the value of the dimensionless inclination angle $\widetilde{\alpha}$. Yet, their respective wakes are qualitatively different. In particular, the wake thickness $h_{w}$ appears to be mostly greater than $h_0$ for large inclinations and lower than $h_0$ for small inclinations. This sizable difference plays a key role in defining the flow regimes. Along its trajectory, a drop indeed absorbs the uniform film laying at its front and releases liquid in its wake. The contribution from the Rayleigh-Taylor instability being negligible (see Fig.~S3 in SM~\cite{Note1}), the change in volume of the drop is $\partial_x V\approx \int_{-R}^{R} \left(h_0-h_{w}(y)\right)\mathrm{d}y$ with $R$ the drop radius. The drop shrinks if $\partial_x V<0$, i.e. if the average thickness left in the wake is less than  $h_0$ as seen for the greater values of the inclination. The structure of the wake is thus key for rationalizing the transition between the two flow regimes.  This numerical observation is confirmed in experiments: as evident from Fig.~\ref{fg:fig4}(b) the wake is thinner than $h_0$ for $\alpha<\alpha_c$ and thicker for $\alpha>\alpha_c$ (see details in SM section I~\cite{Note1}).


We model the variation in thickness across the wake using an approach analogous to that used in Landau-Levich and Bretherton problems. We treat our problem in the polar coordinate system centered on the drop apex [see Fig.~\ref{fg:fig4}(c)]. Focusing on the matching region between the drop and the film, we expect the radial curvature to vary rapidly and dominate the pressure gradient~\cite{Lister:2010}. Consequently, we neglect the advection and gravity terms in the meniscus such that eq.~\eqref{eq:lub_adim} reduces to a radial Landau-Levich equation (see SM section III~\cite{Note1}). Therefore, we treat the wake as a collection of two-dimensional radial Landau-Levich films, where the projected speed $U \cos(\theta)$ is the effective deposition speed. In this framework, we obtain~\cite{Cantat:2013}
\begin{equation}
h_{w}(\theta)\approx 1.34\kappa_d^{-1}\cos\left(\theta\right)^{2/3}\mathrm{Ca}^{2/3}.
\label{eq:hw}
\end{equation}
 Here $\kappa_d$ is the curvature at the edge of the drop, which is assumed to remain close to that of a static pendant drop $\kappa_d\approx 0.28 A/\ell_c^2$ (see SM section III~\cite{Note1}), and $\mathrm{Ca}=\eta U/\gamma$ is the capillary number of the problem. Note that the drop speed $U(t)$ and amplitude $A(t)$ are a priori unknown and depend on the drop initial profile, the film thickness and the inclination of the substrate.
Varying the dimensionless parameters of the problem ($A_0/\ell_c$, $h_0/\ell_c$ and $\widetilde{\alpha}$) we generate a large data set of simulations to assess the validity of our model.

\begin{figure}[tb]
    \begin{center}
        \includegraphics[width=0.9\columnwidth]{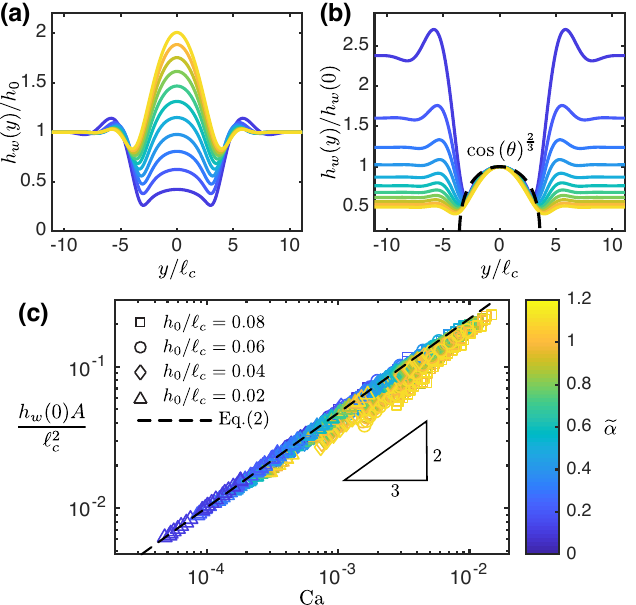}
        \caption{\textbf{(a)} Wake profile $h_w(y)/h_0$ taken $\approx 6\ell_c$ behind a drop~\cite{Note1} for different inclination angles $\widetilde{\alpha}$ [color coded see \textbf{(c)}], $A_0/\ell_c=1.4$ and $h_0/\ell_c=0.04$. \textbf{(b)} Same data rescaled by the central thickness $h_w(0)$. The black dashed line derives from eq.~\eqref{eq:hw}: $h_w(y)/h(0)=\cos(\theta)^{2/3}=\cos(\arcsin(y/R))^{2/3}$. \textbf{(c)} Dimensionless thickness in the center of the wake $h_w(0)A/\ell_c^2$ as a function of the capillary number $\mathrm{Ca}$ for our $176$ simulations. The color codes $\widetilde{\alpha}$, the symbols code $h_0/\ell_c$ and the black dashed line corresponds to eq.~\eqref{eq:hw}: $y=1.34/0.28 \, x^{2/3}$.
        } 
        \label{fg:wake}
    \end{center}
\end{figure}
In Fig.~\ref{fg:wake}(a) we plot the wake profile in the transverse direction $h_w(y)$ for a given drop and film at different inclinations angles [the wake is quasi-invariant in the $x$ direction, see Fig.~\ref{fg:fig4}(a)]. We first focus on the angular dependence by rescaling the data by $h_w(y=0)$. As shown in Fig.~\ref{fg:wake}(b), the profiles collapse in the central region of the wake defined as $-R<y<R$ with $R$ the drop radius. The resulting master curve matches our theoretical prediction $h_w(\theta)/h_w(0)=\cos\left(\theta\right)^{2/3}$ with no fitting parameter [see eq.~\eqref{eq:hw}]. We then compare our prediction for $h_w(0)$ to data from all our simulations in Fig.~\ref{fg:wake}(c). Note that each simulation provides multiple data points as $A$ and $U$ are both function of time and thus vary over the course of a simulation. The resulting collapse and overall favorable agreement with eq.~\eqref{eq:hw} confirms the validity of our approach. Note that the agreement becomes less favorable when $\widetilde{\alpha}$ and $h_0/\ell_c$ increase, a result consistent with the approximations made in our model (negligible advection in the meniscus and static pendant drop shape, see SM section III~\cite{Note1}).


Using eq.~\eqref{eq:hw}, we evaluate the amount of liquid deposited in the wake  $\int_{-R}^{R} h_{w}(y)\mathrm{d}y= R h_{w}(0)\int_{-\pi/2}^{\pi/2}\cos\left(\theta\right)^{5/3}\mathrm{d}\theta$ and obtain the drop growth rate 
 \begin{equation}
 \partial_x V\approx R\left(2 h_0-7.91\ell_c^2\mathrm{Ca}^{2/3}/A\right).
 \label{eq:dvx}
 \end{equation} 
Next, we derive an expression for $\mathrm{Ca}$ in order to close the problem.

\begin{figure}[tb]
    \begin{center}
        \includegraphics[width=0.9\columnwidth]{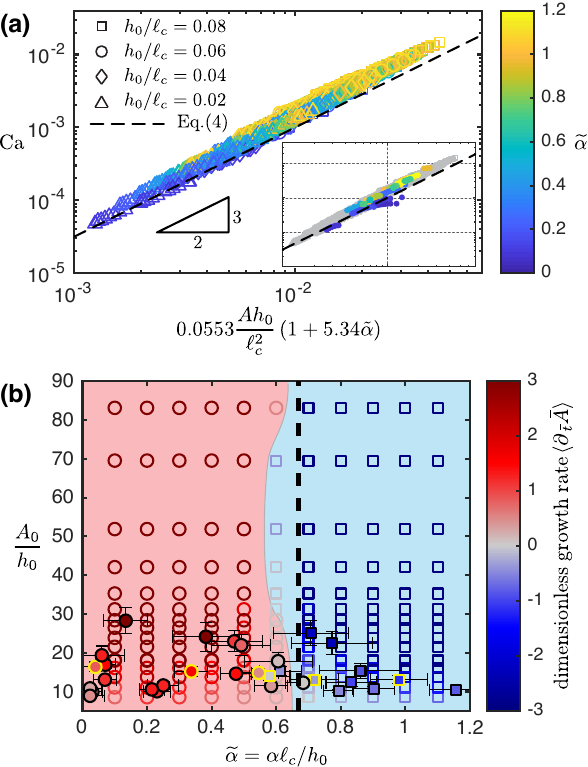}
        \caption{\textbf{(a)} Drop dimensionless speed $\mathrm{Ca}$ compared to our theory [eq.~\eqref{eq:drop_speed}]. Our $176$ simulations are shown, the color codes $\widetilde{\alpha}$, the symbols code $h_0/\ell_c$ and the black dashed line is our prediction $y=x^{3/2}$. Inset: Same plot including experimental data ($0.03<h_0/\ell_c<0.1$). The numerics are drawn in light grey for clarity.     
         \textbf{(b)} State diagram for the drop growth-decay. Symbols are colored according to the dimensionless growth rate averaged over the experiment/simulation $\langle\partial_{\,\bar{t}} \bar{A}\rangle$, circles represent growth ($\langle\partial_{\,\bar{t}} \bar{A}\rangle>0$) while squares represent decay ($\langle\partial_{\,\bar{t}} \bar{A}\rangle<0$). Experimental results are shown as filled symbols while numerical results are open symbols.
         The experiments of Fig.~\ref{fg:fig3} are circled in yellow. The background color is a guide to the eye and the dashed black line is our theory $\widetilde{\alpha}_c\approx 0.67$.
} 
        \label{fg:fig5}
    \end{center}
\end{figure}

To obtain the drop speed, we perform a force balance on the drop~\cite{Aussillous:2002}. The force driving the motion of the drop derives from the change in gravitational energy $E_p=\rho g V z_c$, with $z_c$ the altitude of the drop center of mass. Defining $F_g=-\partial_x E_p$ we find $F_g\approx \rho g V \alpha + 2\rho g z_c R h_0 - 7.91\gamma R z_c \mathrm{Ca}^{2/3}/A$. While the first term in the expression of $F_g$ is conventional, the other two terms originate from the change of volume of the drop $\partial_x V$ in eq.~\eqref{eq:dvx}. The motion of the drop is resisted by viscous stresses in the film.
The flow being significant only around the drop and the meniscus being the thinnest part of that region, we anticipate the meniscus to be the main source of dissipation. 
The corresponding viscous force per unit length is $f_{v}(\theta)=4.94\gamma \left(\mathrm{Ca}\cos\theta\right)^{2/3}$~\cite{Cantat:2013}. Integrating along the drop contour, the total friction force is $\mathbf{F_{v}}=\int_{-\pi/2}^{\pi/2}f_{v}(\theta)\left(\cos\theta \mathbf{e_x}+\sin\theta \mathbf{e_y}\right) R\mathrm{d}\theta\approx 8.31\gamma R \mathrm{Ca}^{2/3}\mathbf{e_x}$. Assuming that the drop shape remains close to that of a static pendant drop we have $z_c\approx 0.29A$, $R\approx 3.58\ell_c$ and $V\approx 0.89 A R^2$ (see SM section III~\cite{Note1} for the derivation of all the prefactors).  Balancing $F_g$ and $F_v=\mathbf{F_{v}}.\mathbf{e_x}$, we obtain:
\begin{equation}
\mathrm{Ca}^{2/3}= 0.0553 \frac{A h_0}{\ell_c^2}\left(1+ 5.34 \widetilde{\alpha}\right).
\label{eq:drop_speed}
\end{equation}
In Fig.~\ref{fg:fig5}(a) we compare the drop speed obtained in simulations with eq.~\eqref{eq:drop_speed} and find favorable agreement without any fitting parameter. Likewise, we show in Inset of Fig.~\ref{fg:fig5}(a) that eq.~\eqref{eq:drop_speed} also captures our experiments. Note that, the agreement becomes less favorable when $\widetilde{\alpha}$ and $h_0/\ell_c$ increase, as expected from the deterioration of our model's assumptions (static pendant drop shape and negligible advection in the meniscus, see SM section III~\cite{Note1}).


We now leverage our results and combine eq.~\eqref{eq:dvx} and eq.~\eqref{eq:drop_speed} to derive the drop growth rate $\partial_x V$ and subsequently integrate this expression over time to obtain the drop dimensionless amplitude 

\begin{equation}
\label{eq:A}
\frac{\bar{A}(\bar{t})}{\bar{A}_0}= \left(1-\frac{1}{2}\f(\widetilde{\alpha})\bar{A}_0^{1/2} \bar{t}\right)^{-2}
\end{equation}  
 with $\f(\widetilde{\alpha})=(0.0065 -0.0097\widetilde{\alpha})(1+5.34\widetilde{\alpha})^{3/2}$ (see SM section III~\cite{Note1}). In Fig.~S5(a) we show that eq.~\eqref{eq:A} compares favorably with experiments without fitting parameters. The value of critical inclination is obtained solving for the root of $\f$, yielding $\widetilde{\alpha}_c \approx 0.67$. 
In Fig.~\ref{fg:fig5}(b) we show the combined experimental and numerical state diagram for the drop dynamics, where the two flow regimes are apparent. As predicted by our model, the transition occurs at a roughly constant critical angle $\widetilde{\alpha}_c\approx 0.6$ in good agreement with our estimate.


        In summary, using experiments and numerical simulations, we have revealed a transition from growth to decay for pendant drops sliding under slightly inclined pre-wet substrates. This transition, which occurs at a surprisingly low angle, is governed by the amount of fluid left in the wake of the drop. As the inclination angle increases, the drop becomes too fast to grow and its volume is slowly depleted. We have rationalized this complex non-linear problem with an analytically tractable Landau-Levich model that accurately predicts the drop dynamics in the two regimes, in spite of the approximations introduced in its derivation.
    Note that on longer time scales the Rayleigh-Taylor instability will eventually influence the dynamics of drops that do not drip. Although this situation is beyond the scope of the present study, preliminary results indicate that the wake forms lenses that are later absorbed by the drop (see SM section II \cite{Note1} and Movie.~S3). Yet, no dripping is observed which suggest that the critical angle we have introduced remains accurate.
As such, our results could find application in dripping prevention for drops directly deposited on substrates, e.g. in coating and printing technologies~\cite{Kumar:2015}.
Additionally, our results could be extended to control and transport pendant drops via carefully crafted substrate topography. Finally, our analysis could be generalized to model the dynamics of sliding liquid plugs in pre-wetted channels~\cite{Bico:2002} and sliding liquid bridges between pre-wetted substrates~\cite{Reyssat:2014,Balestra:2018b}.\\

\begin{acknowledgments}
\noindent\textbf{Acknowledgments:} We thank P. Bourrianne for measuring the silicone oil viscosity. E.J.-P. was partially  supported by NSF through the Princeton University’s Materials Research Science and Engineering Center DMR-1420541. P.G.L. acknowledges the Swiss National Science Foundation under grant 200021-178971.
\end{acknowledgments}


\begin{thebibliography}{30}%
\makeatletter
\providecommand \@ifxundefined [1]{%
 \@ifx{#1\undefined}
}%
\providecommand \@ifnum [1]{%
 \ifnum #1\expandafter \@firstoftwo
 \else \expandafter \@secondoftwo
 \fi
}%
\providecommand \@ifx [1]{%
 \ifx #1\expandafter \@firstoftwo
 \else \expandafter \@secondoftwo
 \fi
}%
\providecommand \natexlab [1]{#1}%
\providecommand \enquote  [1]{``#1''}%
\providecommand \bibnamefont  [1]{#1}%
\providecommand \bibfnamefont [1]{#1}%
\providecommand \citenamefont [1]{#1}%
\providecommand \href@noop [0]{\@secondoftwo}%
\providecommand \href [0]{\begingroup \@sanitize@url \@href}%
\providecommand \@href[1]{\@@startlink{#1}\@@href}%
\providecommand \@@href[1]{\endgroup#1\@@endlink}%
\providecommand \@sanitize@url [0]{\catcode `\\12\catcode `\$12\catcode
  `\&12\catcode `\#12\catcode `\^12\catcode `\_12\catcode `\%12\relax}%
\providecommand \@@startlink[1]{}%
\providecommand \@@endlink[0]{}%
\providecommand \url  [0]{\begingroup\@sanitize@url \@url }%
\providecommand \@url [1]{\endgroup\@href {#1}{\urlprefix }}%
\providecommand \urlprefix  [0]{URL }%
\providecommand \Eprint [0]{\href }%
\providecommand \doibase [0]{https://doi.org/}%
\providecommand \selectlanguage [0]{\@gobble}%
\providecommand \bibinfo  [0]{\@secondoftwo}%
\providecommand \bibfield  [0]{\@secondoftwo}%
\providecommand \translation [1]{[#1]}%
\providecommand \BibitemOpen [0]{}%
\providecommand \bibitemStop [0]{}%
\providecommand \bibitemNoStop [0]{.\EOS\space}%
\providecommand \EOS [0]{\spacefactor3000\relax}%
\providecommand \BibitemShut  [1]{\csname bibitem#1\endcsname}%
\let\auto@bib@innerbib\@empty
\bibitem [{\citenamefont {Yiantsios}\ and\ \citenamefont
  {Higgins}(1989)}]{Yiantsios:1989}%
  \BibitemOpen
  \bibfield  {author} {\bibinfo {author} {\bibfnamefont {S.~G.}\ \bibnamefont
  {Yiantsios}}\ and\ \bibinfo {author} {\bibfnamefont {B.~G.}\ \bibnamefont
  {Higgins}},\ }\bibfield  {title} {\bibinfo {title} {Rayleigh–taylor
  instability in thin viscous films},\ }\href@noop {} {\bibfield  {journal}
  {\bibinfo  {journal} {Physics of Fluids A: Fluid Dynamics}\ }\textbf
  {\bibinfo {volume} {1}},\ \bibinfo {pages} {1484} (\bibinfo {year}
  {1989})}\BibitemShut {NoStop}%
\bibitem [{\citenamefont {Fermigier}\ \emph {et~al.}(1992)\citenamefont
  {Fermigier}, \citenamefont {Limat}, \citenamefont {Wesfreid}, \citenamefont
  {Boudinet},\ and\ \citenamefont {Quilliet}}]{Fermigier:1992}%
  \BibitemOpen
  \bibfield  {author} {\bibinfo {author} {\bibfnamefont {M.}~\bibnamefont
  {Fermigier}}, \bibinfo {author} {\bibfnamefont {L.}~\bibnamefont {Limat}},
  \bibinfo {author} {\bibfnamefont {J.~E.}\ \bibnamefont {Wesfreid}}, \bibinfo
  {author} {\bibfnamefont {P.}~\bibnamefont {Boudinet}},\ and\ \bibinfo
  {author} {\bibfnamefont {C.}~\bibnamefont {Quilliet}},\ }\bibfield  {title}
  {\bibinfo {title} {Two-dimensional patterns in rayleigh-taylor instability of
  a thin layer},\ }\href@noop {} {\bibfield  {journal} {\bibinfo  {journal}
  {Journal of Fluid Mechanics}\ }\textbf {\bibinfo {volume} {236}},\ \bibinfo
  {pages} {349–383} (\bibinfo {year} {1992})}\BibitemShut {NoStop}%
\bibitem [{\citenamefont {Weitz}\ \emph {et~al.}(2008)\citenamefont {Weitz},
  \citenamefont {Harnau}, \citenamefont {Rauschenbach}, \citenamefont
  {Burghard},\ and\ \citenamefont {Kern}}]{Weitz:2008}%
  \BibitemOpen
  \bibfield  {author} {\bibinfo {author} {\bibfnamefont {R.~T.}\ \bibnamefont
  {Weitz}}, \bibinfo {author} {\bibfnamefont {L.}~\bibnamefont {Harnau}},
  \bibinfo {author} {\bibfnamefont {S.}~\bibnamefont {Rauschenbach}}, \bibinfo
  {author} {\bibfnamefont {M.}~\bibnamefont {Burghard}},\ and\ \bibinfo
  {author} {\bibfnamefont {K.}~\bibnamefont {Kern}},\ }\bibfield  {title}
  {\bibinfo {title} {Polymer nanofibers via nozzle-free centrifugal spinning},\
  }\href@noop {} {\bibfield  {journal} {\bibinfo  {journal} {Nano Letters}\
  }\textbf {\bibinfo {volume} {8}},\ \bibinfo {pages} {1187} (\bibinfo {year}
  {2008})}\BibitemShut {NoStop}%
\bibitem [{\citenamefont {Marthelot}\ \emph {et~al.}(2018)\citenamefont
  {Marthelot}, \citenamefont {Strong}, \citenamefont {Reis},\ and\
  \citenamefont {Brun}}]{Marthelot:2018}%
  \BibitemOpen
  \bibfield  {author} {\bibinfo {author} {\bibfnamefont {J.}~\bibnamefont
  {Marthelot}}, \bibinfo {author} {\bibfnamefont {E.~F.}\ \bibnamefont
  {Strong}}, \bibinfo {author} {\bibfnamefont {P.~M.}\ \bibnamefont {Reis}},\
  and\ \bibinfo {author} {\bibfnamefont {P.-T.}\ \bibnamefont {Brun}},\
  }\bibfield  {title} {\bibinfo {title} {Designing soft materials with
  interfacial instabilities in liquid films},\ }\href@noop {} {\bibfield
  {journal} {\bibinfo  {journal} {Nature Communications}\ }\textbf {\bibinfo
  {volume} {9}},\ \bibinfo {pages} {4477} (\bibinfo {year} {2018})}\BibitemShut
  {NoStop}%
\bibitem [{\citenamefont {Jambon-Puillet}\ \emph {et~al.}(2021)\citenamefont
  {Jambon-Puillet}, \citenamefont {Royer~Pi{\'e}chaud},\ and\ \citenamefont
  {Brun}}]{Jambon:2021}%
  \BibitemOpen
  \bibfield  {author} {\bibinfo {author} {\bibfnamefont {E.}~\bibnamefont
  {Jambon-Puillet}}, \bibinfo {author} {\bibfnamefont {M.}~\bibnamefont
  {Royer~Pi{\'e}chaud}},\ and\ \bibinfo {author} {\bibfnamefont {P.-T.}\
  \bibnamefont {Brun}},\ }\bibfield  {title} {\bibinfo {title} {Elastic
  amplification of the rayleigh-taylor instability in solidifying melts},\
  }\href {https://www.pnas.org/content/118/10/e2020701118} {\bibfield
  {journal} {\bibinfo  {journal} {Proceedings of the National Academy of
  Sciences}\ }\textbf {\bibinfo {volume} {118}} (\bibinfo {year}
  {2021})}\BibitemShut {NoStop}%
\bibitem [{\citenamefont {Ribe}\ \emph {et~al.}(2007)\citenamefont {Ribe},
  \citenamefont {Davaille},\ and\ \citenamefont {Christensen}}]{Ribe:2007}%
  \BibitemOpen
  \bibfield  {author} {\bibinfo {author} {\bibfnamefont {N.}~\bibnamefont
  {Ribe}}, \bibinfo {author} {\bibfnamefont {A.}~\bibnamefont {Davaille}},\
  and\ \bibinfo {author} {\bibfnamefont {U.}~\bibnamefont {Christensen}},\
  }\bibinfo {title} {Fluid dynamics of mantle plumes},\ in\ \href@noop {}
  {\emph {\bibinfo {booktitle} {Mantle Plumes: A Multidisciplinary
  Approach}}},\ \bibinfo {editor} {edited by\ \bibinfo {editor} {\bibfnamefont
  {J.~R.~R.}\ \bibnamefont {Ritter}}\ and\ \bibinfo {editor} {\bibfnamefont
  {U.~R.}\ \bibnamefont {Christensen}}}\ (\bibinfo  {publisher} {Springer
  Berlin Heidelberg},\ \bibinfo {address} {Berlin, Heidelberg},\ \bibinfo
  {year} {2007})\ pp.\ \bibinfo {pages} {1--48}\BibitemShut {NoStop}%
\bibitem [{\citenamefont {Camporeale}\ and\ \citenamefont
  {Ridolfi}(2012)}]{camporeale2012hydrodynamic}%
  \BibitemOpen
  \bibfield  {author} {\bibinfo {author} {\bibfnamefont {C.}~\bibnamefont
  {Camporeale}}\ and\ \bibinfo {author} {\bibfnamefont {L.}~\bibnamefont
  {Ridolfi}},\ }\bibfield  {title} {\bibinfo {title} {Hydrodynamic-driven
  stability analysis of morphological patterns on stalactites and implications
  for cave paleoflow reconstructions},\ }\href@noop {} {\bibfield  {journal}
  {\bibinfo  {journal} {Phys. Rev. Lett.}\ }\textbf {\bibinfo {volume} {108}},\
  \bibinfo {pages} {238501} (\bibinfo {year} {2012})}\BibitemShut {NoStop}%
\bibitem [{\citenamefont {Dutta}\ \emph {et~al.}(2016)\citenamefont {Dutta},
  \citenamefont {Baruah},\ and\ \citenamefont {Mandal}}]{Dutta:2016}%
  \BibitemOpen
  \bibfield  {author} {\bibinfo {author} {\bibfnamefont {U.}~\bibnamefont
  {Dutta}}, \bibinfo {author} {\bibfnamefont {A.}~\bibnamefont {Baruah}},\ and\
  \bibinfo {author} {\bibfnamefont {N.}~\bibnamefont {Mandal}},\ }\bibfield
  {title} {\bibinfo {title} {{Role of source-layer tilts in the axi-asymmetric
  growth of diapirs triggered by a Rayleigh–Taylor instability}},\
  }\href@noop {} {\bibfield  {journal} {\bibinfo  {journal} {Geophysical
  Journal International}\ }\textbf {\bibinfo {volume} {206}},\ \bibinfo {pages}
  {1814} (\bibinfo {year} {2016})}\BibitemShut {NoStop}%
\bibitem [{\citenamefont {Weinstein}\ and\ \citenamefont
  {Ruschak}(2004)}]{Weinstein:2004}%
  \BibitemOpen
  \bibfield  {author} {\bibinfo {author} {\bibfnamefont {S.~J.}\ \bibnamefont
  {Weinstein}}\ and\ \bibinfo {author} {\bibfnamefont {K.~J.}\ \bibnamefont
  {Ruschak}},\ }\bibfield  {title} {\bibinfo {title} {Coating flows},\
  }\href@noop {} {\bibfield  {journal} {\bibinfo  {journal} {Annual Review of
  Fluid Mechanics}\ }\textbf {\bibinfo {volume} {36}},\ \bibinfo {pages} {29}
  (\bibinfo {year} {2004})}\BibitemShut {NoStop}%
\bibitem [{\citenamefont {Kaita}\ \emph {et~al.}(2010)\citenamefont {Kaita},
  \citenamefont {Berzak}, \citenamefont {Boyle}, \citenamefont {Gray},
  \citenamefont {Granstedt}, \citenamefont {Hammett}, \citenamefont {Jacobson},
  \citenamefont {Jones}, \citenamefont {Kozub}, \citenamefont {Kugel},
  \citenamefont {Leblanc}, \citenamefont {Logan}, \citenamefont {Lucia},
  \citenamefont {Lundberg}, \citenamefont {Majeski}, \citenamefont {Mansfield},
  \citenamefont {Menard}, \citenamefont {Spaleta}, \citenamefont {Strickler},
  \citenamefont {Timberlake}, \citenamefont {Yoo}, \citenamefont {Zakharov},
  \citenamefont {Maingi}, \citenamefont {Soukhanovskii}, \citenamefont
  {Tritz},\ and\ \citenamefont {Gershman}}]{Kaita:2010}%
  \BibitemOpen
  \bibfield  {author} {\bibinfo {author} {\bibfnamefont {R.}~\bibnamefont
  {Kaita}}, \bibinfo {author} {\bibfnamefont {L.}~\bibnamefont {Berzak}},
  \bibinfo {author} {\bibfnamefont {D.}~\bibnamefont {Boyle}}, \bibinfo
  {author} {\bibfnamefont {T.}~\bibnamefont {Gray}}, \bibinfo {author}
  {\bibfnamefont {E.}~\bibnamefont {Granstedt}}, \bibinfo {author}
  {\bibfnamefont {G.}~\bibnamefont {Hammett}}, \bibinfo {author} {\bibfnamefont
  {C.~M.}\ \bibnamefont {Jacobson}}, \bibinfo {author} {\bibfnamefont
  {A.}~\bibnamefont {Jones}}, \bibinfo {author} {\bibfnamefont
  {T.}~\bibnamefont {Kozub}}, \bibinfo {author} {\bibfnamefont
  {H.}~\bibnamefont {Kugel}}, \bibinfo {author} {\bibfnamefont
  {B.}~\bibnamefont {Leblanc}}, \bibinfo {author} {\bibfnamefont
  {N.}~\bibnamefont {Logan}}, \bibinfo {author} {\bibfnamefont
  {M.}~\bibnamefont {Lucia}}, \bibinfo {author} {\bibfnamefont
  {D.}~\bibnamefont {Lundberg}}, \bibinfo {author} {\bibfnamefont
  {R.}~\bibnamefont {Majeski}}, \bibinfo {author} {\bibfnamefont
  {D.}~\bibnamefont {Mansfield}}, \bibinfo {author} {\bibfnamefont
  {J.}~\bibnamefont {Menard}}, \bibinfo {author} {\bibfnamefont
  {J.}~\bibnamefont {Spaleta}}, \bibinfo {author} {\bibfnamefont
  {T.}~\bibnamefont {Strickler}}, \bibinfo {author} {\bibfnamefont
  {J.}~\bibnamefont {Timberlake}}, \bibinfo {author} {\bibfnamefont
  {J.}~\bibnamefont {Yoo}}, \bibinfo {author} {\bibfnamefont {L.}~\bibnamefont
  {Zakharov}}, \bibinfo {author} {\bibfnamefont {R.}~\bibnamefont {Maingi}},
  \bibinfo {author} {\bibfnamefont {V.}~\bibnamefont {Soukhanovskii}}, \bibinfo
  {author} {\bibfnamefont {K.}~\bibnamefont {Tritz}},\ and\ \bibinfo {author}
  {\bibfnamefont {S.}~\bibnamefont {Gershman}},\ }\bibfield  {title} {\bibinfo
  {title} {Experiments with liquid metal walls: Status of the lithium tokamak
  experiment},\ }\href@noop {} {\bibfield  {journal} {\bibinfo  {journal}
  {Fusion Engineering and Design}\ }\textbf {\bibinfo {volume} {85}},\ \bibinfo
  {pages} {874 } (\bibinfo {year} {2010})}\BibitemShut {NoStop}%
\bibitem [{\citenamefont {van Eden}\ \emph {et~al.}(2017)\citenamefont {van
  Eden}, \citenamefont {Kvon}, \citenamefont {van~de Sanden},\ and\
  \citenamefont {Morgan}}]{vanEden:2017}%
  \BibitemOpen
  \bibfield  {author} {\bibinfo {author} {\bibfnamefont {G.~G.}\ \bibnamefont
  {van Eden}}, \bibinfo {author} {\bibfnamefont {V.}~\bibnamefont {Kvon}},
  \bibinfo {author} {\bibfnamefont {M.~C.~M.}\ \bibnamefont {van~de Sanden}},\
  and\ \bibinfo {author} {\bibfnamefont {T.~W.}\ \bibnamefont {Morgan}},\
  }\bibfield  {title} {\bibinfo {title} {Oscillatory vapour shielding of liquid
  metal walls in nuclear fusion devices},\ }\href@noop {} {\bibfield  {journal}
  {\bibinfo  {journal} {Nature Communications}\ }\textbf {\bibinfo {volume}
  {8}},\ \bibinfo {pages} {192} (\bibinfo {year} {2017})}\BibitemShut {NoStop}%
\bibitem [{\citenamefont {Limat}\ \emph {et~al.}(1992)\citenamefont {Limat},
  \citenamefont {Jenffer}, \citenamefont {Dagens}, \citenamefont {Touron},
  \citenamefont {Fermigier},\ and\ \citenamefont {Wesfreid}}]{Limat:1992}%
  \BibitemOpen
  \bibfield  {author} {\bibinfo {author} {\bibfnamefont {L.}~\bibnamefont
  {Limat}}, \bibinfo {author} {\bibfnamefont {P.}~\bibnamefont {Jenffer}},
  \bibinfo {author} {\bibfnamefont {B.}~\bibnamefont {Dagens}}, \bibinfo
  {author} {\bibfnamefont {E.}~\bibnamefont {Touron}}, \bibinfo {author}
  {\bibfnamefont {M.}~\bibnamefont {Fermigier}},\ and\ \bibinfo {author}
  {\bibfnamefont {J.}~\bibnamefont {Wesfreid}},\ }\bibfield  {title} {\bibinfo
  {title} {Gravitational instabilities of thin liquid layers: dynamics of
  pattern selection},\ }\href@noop {} {\bibfield  {journal} {\bibinfo
  {journal} {Physica D: Nonlinear Phenomena}\ }\textbf {\bibinfo {volume}
  {61}},\ \bibinfo {pages} {166 } (\bibinfo {year} {1992})}\BibitemShut
  {NoStop}%
\bibitem [{\citenamefont {Yoshikawa}\ \emph {et~al.}(2019)\citenamefont
  {Yoshikawa}, \citenamefont {Mathis}, \citenamefont {Satoh},\ and\
  \citenamefont {Tasaka}}]{Yoshikawa:2019}%
  \BibitemOpen
  \bibfield  {author} {\bibinfo {author} {\bibfnamefont {H.~N.}\ \bibnamefont
  {Yoshikawa}}, \bibinfo {author} {\bibfnamefont {C.}~\bibnamefont {Mathis}},
  \bibinfo {author} {\bibfnamefont {S.}~\bibnamefont {Satoh}},\ and\ \bibinfo
  {author} {\bibfnamefont {Y.}~\bibnamefont {Tasaka}},\ }\bibfield  {title}
  {\bibinfo {title} {Inwardly rotating spirals in a nonoscillatory medium},\
  }\href@noop {} {\bibfield  {journal} {\bibinfo  {journal} {Phys. Rev. Lett.}\
  }\textbf {\bibinfo {volume} {122}},\ \bibinfo {pages} {014502} (\bibinfo
  {year} {2019})}\BibitemShut {NoStop}%
\bibitem [{\citenamefont {Lerisson}\ \emph {et~al.}(2020)\citenamefont
  {Lerisson}, \citenamefont {Ledda}, \citenamefont {Balestra},\ and\
  \citenamefont {Gallaire}}]{Lerisson:2020}%
  \BibitemOpen
  \bibfield  {author} {\bibinfo {author} {\bibfnamefont {G.}~\bibnamefont
  {Lerisson}}, \bibinfo {author} {\bibfnamefont {P.~G.}\ \bibnamefont {Ledda}},
  \bibinfo {author} {\bibfnamefont {G.}~\bibnamefont {Balestra}},\ and\
  \bibinfo {author} {\bibfnamefont {F.}~\bibnamefont {Gallaire}},\ }\bibfield
  {title} {\bibinfo {title} {Instability of a thin viscous film flowing under
  an inclined substrate: steady patterns},\ }\href@noop {} {\bibfield
  {journal} {\bibinfo  {journal} {Journal of Fluid Mechanics}\ }\textbf
  {\bibinfo {volume} {898}},\ \bibinfo {pages} {A6} (\bibinfo {year}
  {2020})}\BibitemShut {NoStop}%
\bibitem [{\citenamefont {Burgess}\ \emph {et~al.}(2001)\citenamefont
  {Burgess}, \citenamefont {Juel}, \citenamefont {McCormick}, \citenamefont
  {Swift},\ and\ \citenamefont {Swinney}}]{Burgess:2001}%
  \BibitemOpen
  \bibfield  {author} {\bibinfo {author} {\bibfnamefont {J.~M.}\ \bibnamefont
  {Burgess}}, \bibinfo {author} {\bibfnamefont {A.}~\bibnamefont {Juel}},
  \bibinfo {author} {\bibfnamefont {W.~D.}\ \bibnamefont {McCormick}}, \bibinfo
  {author} {\bibfnamefont {J.~B.}\ \bibnamefont {Swift}},\ and\ \bibinfo
  {author} {\bibfnamefont {H.~L.}\ \bibnamefont {Swinney}},\ }\bibfield
  {title} {\bibinfo {title} {Suppression of dripping from a ceiling},\
  }\href@noop {} {\bibfield  {journal} {\bibinfo  {journal} {Phys. Rev. Lett.}\
  }\textbf {\bibinfo {volume} {86}},\ \bibinfo {pages} {1203} (\bibinfo {year}
  {2001})}\BibitemShut {NoStop}%
\bibitem [{\citenamefont {Lapuerta}\ \emph {et~al.}(2001)\citenamefont
  {Lapuerta}, \citenamefont {Mancebo},\ and\ \citenamefont
  {Vega}}]{Lapuerta:2001}%
  \BibitemOpen
  \bibfield  {author} {\bibinfo {author} {\bibfnamefont {V.}~\bibnamefont
  {Lapuerta}}, \bibinfo {author} {\bibfnamefont {F.~J.}\ \bibnamefont
  {Mancebo}},\ and\ \bibinfo {author} {\bibfnamefont {J.~M.}\ \bibnamefont
  {Vega}},\ }\bibfield  {title} {\bibinfo {title} {Control of rayleigh-taylor
  instability by vertical vibration in large aspect ratio containers},\
  }\href@noop {} {\bibfield  {journal} {\bibinfo  {journal} {Phys. Rev. E}\
  }\textbf {\bibinfo {volume} {64}},\ \bibinfo {pages} {016318} (\bibinfo
  {year} {2001})}\BibitemShut {NoStop}%
\bibitem [{\citenamefont {Alexeev}\ and\ \citenamefont
  {Oron}(2007)}]{Alexeev:2007}%
  \BibitemOpen
  \bibfield  {author} {\bibinfo {author} {\bibfnamefont {A.}~\bibnamefont
  {Alexeev}}\ and\ \bibinfo {author} {\bibfnamefont {A.}~\bibnamefont {Oron}},\
  }\bibfield  {title} {\bibinfo {title} {Suppression of the rayleigh-taylor
  instability of thin liquid films by the marangoni effect},\ }\href@noop {}
  {\bibfield  {journal} {\bibinfo  {journal} {Physics of Fluids}\ }\textbf
  {\bibinfo {volume} {19}},\ \bibinfo {pages} {082101} (\bibinfo {year}
  {2007})}\BibitemShut {NoStop}%
\bibitem [{\citenamefont {Cimpeanu}\ \emph {et~al.}(2014)\citenamefont
  {Cimpeanu}, \citenamefont {Papageorgiou},\ and\ \citenamefont
  {Petropoulos}}]{Cimpeanu:2014}%
  \BibitemOpen
  \bibfield  {author} {\bibinfo {author} {\bibfnamefont {R.}~\bibnamefont
  {Cimpeanu}}, \bibinfo {author} {\bibfnamefont {D.~T.}\ \bibnamefont
  {Papageorgiou}},\ and\ \bibinfo {author} {\bibfnamefont {P.~G.}\ \bibnamefont
  {Petropoulos}},\ }\bibfield  {title} {\bibinfo {title} {On the control and
  suppression of the rayleigh-taylor instability using electric fields},\
  }\href@noop {} {\bibfield  {journal} {\bibinfo  {journal} {Physics of
  Fluids}\ }\textbf {\bibinfo {volume} {26}},\ \bibinfo {pages} {022105}
  (\bibinfo {year} {2014})}\BibitemShut {NoStop}%
\bibitem [{\citenamefont {Trinh}\ \emph {et~al.}(2014)\citenamefont {Trinh},
  \citenamefont {Kim}, \citenamefont {Hammoud}, \citenamefont {Howell},
  \citenamefont {Chapman},\ and\ \citenamefont {Stone}}]{Trinh:2014}%
  \BibitemOpen
  \bibfield  {author} {\bibinfo {author} {\bibfnamefont {P.~H.}\ \bibnamefont
  {Trinh}}, \bibinfo {author} {\bibfnamefont {H.}~\bibnamefont {Kim}}, \bibinfo
  {author} {\bibfnamefont {N.}~\bibnamefont {Hammoud}}, \bibinfo {author}
  {\bibfnamefont {P.~D.}\ \bibnamefont {Howell}}, \bibinfo {author}
  {\bibfnamefont {S.~J.}\ \bibnamefont {Chapman}},\ and\ \bibinfo {author}
  {\bibfnamefont {H.~A.}\ \bibnamefont {Stone}},\ }\bibfield  {title} {\bibinfo
  {title} {Curvature suppresses the rayleigh-taylor instability},\ }\href@noop
  {} {\bibfield  {journal} {\bibinfo  {journal} {Physics of Fluids}\ }\textbf
  {\bibinfo {volume} {26}},\ \bibinfo {pages} {051704} (\bibinfo {year}
  {2014})}\BibitemShut {NoStop}%
\bibitem [{\citenamefont {Brun}\ \emph {et~al.}(2015)\citenamefont {Brun},
  \citenamefont {Damiano}, \citenamefont {Rieu}, \citenamefont {Balestra},\
  and\ \citenamefont {Gallaire}}]{Brun:2015}%
  \BibitemOpen
  \bibfield  {author} {\bibinfo {author} {\bibfnamefont {P.-T.}\ \bibnamefont
  {Brun}}, \bibinfo {author} {\bibfnamefont {A.}~\bibnamefont {Damiano}},
  \bibinfo {author} {\bibfnamefont {P.}~\bibnamefont {Rieu}}, \bibinfo {author}
  {\bibfnamefont {G.}~\bibnamefont {Balestra}},\ and\ \bibinfo {author}
  {\bibfnamefont {F.}~\bibnamefont {Gallaire}},\ }\bibfield  {title} {\bibinfo
  {title} {Rayleigh-taylor instability under an inclined plane},\ }\href@noop
  {} {\bibfield  {journal} {\bibinfo  {journal} {Physics of Fluids}\ }\textbf
  {\bibinfo {volume} {27}},\ \bibinfo {pages} {084107} (\bibinfo {year}
  {2015})}\BibitemShut {NoStop}%
\bibitem [{\citenamefont {Balestra}\ \emph {et~al.}(2018)\citenamefont
  {Balestra}, \citenamefont {Kofman}, \citenamefont {Brun}, \citenamefont
  {Scheid},\ and\ \citenamefont {Gallaire}}]{Balestra:2018}%
  \BibitemOpen
  \bibfield  {author} {\bibinfo {author} {\bibfnamefont {G.}~\bibnamefont
  {Balestra}}, \bibinfo {author} {\bibfnamefont {N.}~\bibnamefont {Kofman}},
  \bibinfo {author} {\bibfnamefont {P.-T.}\ \bibnamefont {Brun}}, \bibinfo
  {author} {\bibfnamefont {B.}~\bibnamefont {Scheid}},\ and\ \bibinfo {author}
  {\bibfnamefont {F.}~\bibnamefont {Gallaire}},\ }\bibfield  {title} {\bibinfo
  {title} {Three-dimensional rayleigh–taylor instability under a
  unidirectional curved substrate},\ }\href@noop {} {\bibfield  {journal}
  {\bibinfo  {journal} {Journal of Fluid Mechanics}\ }\textbf {\bibinfo
  {volume} {837}},\ \bibinfo {pages} {19–47} (\bibinfo {year}
  {2018})}\BibitemShut {NoStop}%
\bibitem [{\citenamefont {Lister}\ \emph {et~al.}(2010)\citenamefont {Lister},
  \citenamefont {Rallison},\ and\ \citenamefont {Rees}}]{Lister:2010}%
  \BibitemOpen
  \bibfield  {author} {\bibinfo {author} {\bibfnamefont {J.~R.}\ \bibnamefont
  {Lister}}, \bibinfo {author} {\bibfnamefont {J.~M.}\ \bibnamefont
  {Rallison}},\ and\ \bibinfo {author} {\bibfnamefont {S.~J.}\ \bibnamefont
  {Rees}},\ }\bibfield  {title} {\bibinfo {title} {The nonlinear dynamics of
  pendent drops on a thin film coating the underside of a ceiling},\ }\href
  {https://doi.org/10.1017/S002211201000008X} {\bibfield  {journal} {\bibinfo
  {journal} {Journal of Fluid Mechanics}\ }\textbf {\bibinfo {volume} {647}},\
  \bibinfo {pages} {239–264} (\bibinfo {year} {2010})}\BibitemShut {NoStop}%
\bibitem [{Note1()}]{Note1}%
  \BibitemOpen
  \bibinfo {note} {See Supplemental Material at [URL will be inserted by
  publisher] for experimental and numerical methods, additional experimental
  and numerical results and more details about the model and its
  limitations.}\BibitemShut {Stop}%
\bibitem [{\citenamefont {Wilson}(1982)}]{Wilson:1982}%
  \BibitemOpen
  \bibfield  {author} {\bibinfo {author} {\bibfnamefont {S.~D.~R.}\
  \bibnamefont {Wilson}},\ }\bibfield  {title} {\bibinfo {title} {The drag-out
  problem in film coating theory},\ }\href@noop {} {\bibfield  {journal}
  {\bibinfo  {journal} {Journal of Engineering Mathematics}\ }\textbf {\bibinfo
  {volume} {16}},\ \bibinfo {pages} {209} (\bibinfo {year} {1982})}\BibitemShut
  {NoStop}%
\bibitem [{\citenamefont {Cantat}(2013)}]{Cantat:2013}%
  \BibitemOpen
  \bibfield  {author} {\bibinfo {author} {\bibfnamefont {I.}~\bibnamefont
  {Cantat}},\ }\bibfield  {title} {\bibinfo {title} {Liquid meniscus friction
  on a wet plate: Bubbles, lamellae, and foams},\ }\href@noop {} {\bibfield
  {journal} {\bibinfo  {journal} {Physics of Fluids}\ }\textbf {\bibinfo
  {volume} {25}},\ \bibinfo {pages} {031303} (\bibinfo {year}
  {2013})}\BibitemShut {NoStop}%
\bibitem [{\citenamefont {Aussillous}\ and\ \citenamefont
  {Qu{\'{e}}r{\'{e}}}(2002)}]{Aussillous:2002}%
  \BibitemOpen
  \bibfield  {author} {\bibinfo {author} {\bibfnamefont {P.}~\bibnamefont
  {Aussillous}}\ and\ \bibinfo {author} {\bibfnamefont {D.}~\bibnamefont
  {Qu{\'{e}}r{\'{e}}}},\ }\bibfield  {title} {\bibinfo {title} {Bubbles
  creeping in a viscous liquid along a slightly inclined plane},\ }\href@noop
  {} {\bibfield  {journal} {\bibinfo  {journal} {Europhysics Letters ({EPL})}\
  }\textbf {\bibinfo {volume} {59}},\ \bibinfo {pages} {370} (\bibinfo {year}
  {2002})}\BibitemShut {NoStop}%
\bibitem [{\citenamefont {Kumar}(2015)}]{Kumar:2015}%
  \BibitemOpen
  \bibfield  {author} {\bibinfo {author} {\bibfnamefont {S.}~\bibnamefont
  {Kumar}},\ }\bibfield  {title} {\bibinfo {title} {Liquid transfer in printing
  processes: Liquid bridges with moving contact lines},\ }\href@noop {}
  {\bibfield  {journal} {\bibinfo  {journal} {Annual Review of Fluid
  Mechanics}\ }\textbf {\bibinfo {volume} {47}},\ \bibinfo {pages} {67}
  (\bibinfo {year} {2015})}\BibitemShut {NoStop}%
\bibitem [{\citenamefont {Bico}\ and\ \citenamefont
  {Quéré}(2002)}]{Bico:2002}%
  \BibitemOpen
  \bibfield  {author} {\bibinfo {author} {\bibfnamefont {J.}~\bibnamefont
  {Bico}}\ and\ \bibinfo {author} {\bibfnamefont {D.}~\bibnamefont {Quéré}},\
  }\bibfield  {title} {\bibinfo {title} {Self-propelling slugs},\ }\href@noop
  {} {\bibfield  {journal} {\bibinfo  {journal} {Journal of Fluid Mechanics}\
  }\textbf {\bibinfo {volume} {467}},\ \bibinfo {pages} {101–127} (\bibinfo
  {year} {2002})}\BibitemShut {NoStop}%
\bibitem [{\citenamefont {Reyssat}(2014)}]{Reyssat:2014}%
  \BibitemOpen
  \bibfield  {author} {\bibinfo {author} {\bibfnamefont {E.}~\bibnamefont
  {Reyssat}},\ }\bibfield  {title} {\bibinfo {title} {Drops and bubbles in
  wedges},\ }\href@noop {} {\bibfield  {journal} {\bibinfo  {journal} {Journal
  of Fluid Mechanics}\ }\textbf {\bibinfo {volume} {748}},\ \bibinfo {pages}
  {641–662} (\bibinfo {year} {2014})}\BibitemShut {NoStop}%
\bibitem [{\citenamefont {Balestra}(2018)}]{Balestra:2018b}%
  \BibitemOpen
  \bibfield  {author} {\bibinfo {author} {\bibfnamefont {G.~M.~N.}\
  \bibnamefont {Balestra}},\ }\emph {\bibinfo {title} {Pattern formation in
  thin liquid films: from coating-flow instabilities to microfluidic
  droplets}},\ \href@noop {} {Ph.D. thesis},\ \bibinfo {address} {Lausanne}
  (\bibinfo {year} {2018})\BibitemShut {NoStop}%
\end{thebibliography}
%

\end{document}